\documentclass{article}

\PassOptionsToPackage{numbers, compress}{natbib}

 \usepackage[final]{nips_2018}

\usepackage[utf8]{inputenc} 
\usepackage[T1]{fontenc}    
\usepackage{hyperref}       
\usepackage{url}            
\usepackage{booktabs}       
\usepackage{amsfonts}       
\usepackage{nicefrac}       
\usepackage{microtype}      
\usepackage{graphicx}       

\title{Fake News Detection on Social Media using Geometric Deep Learning}

\author{
  Federico Monti$^{1,2}$ \And Fabrizio Frasca$^{1,2}$ \And Davide Eynard$^{1,2}$ \And Damon Mannion$^{1,2}$ \And Michael M. Bronstein$^{1,2,3}$\vspace{4mm}\\
  \begin{tabular}{ c c c  c c }
 $^1$Fabula AI & ~~~ & $^2$USI Lugano & ~~~ & $^3$Imperial College \\ 
 United Kingdom & & Switzerland & & United Kingdom \\  
\end{tabular}
\\
}

\usepackage[utf8]{inputenc}
\usepackage{pgfplots}

\newlength\figureheight
\newlength\figurewidth

\begin{document}

\maketitle

\begin{abstract}

Social media are nowadays one of the main news sources for millions of people around the globe due to their low cost, easy access, and rapid dissemination. This however comes at the cost of dubious trustworthiness and significant risk of exposure to `fake news',  intentionally written to mislead the readers. 
%
Automatically detecting fake news poses challenges that defy existing content-based analysis approaches. One of the main reasons is that  often the interpretation of the news requires the knowledge of political or social context or `common sense', which current natural language processing algorithms are still missing. 
Recent studies have empirically shown that fake and real news spread differently on social media, forming propagation patterns that could be harnessed for the automatic fake news detection. Propagation-based approaches  have multiple advantages compared to their content-based counterparts, among which is language independence and better resilience to adversarial attacks. 
In this paper, we show a novel automatic fake news detection model based on  geometric deep learning.  
%
The underlying core algorithms are a generalization of classical convolutional neural networks to graphs, allowing the fusion of  heterogeneous data such as content, user profile and activity, social graph, and news propagation. 
Our model was trained and tested on news stories, verified by professional fact-checking organizations, that were spread on Twitter. Our experiments indicate that social network structure and propagation are important features allowing highly accurate (92.7\% ROC AUC) fake news detection. Second, we observe that fake news can be reliably detected at an early stage, after just a few hours of propagation. Third, we test the aging of our model on training and testing data separated in time. 
Our results point to the promise of propagation-based approaches for fake news detection as an alternative or complementary strategy to content-based approaches.

\end{abstract}

\pagebreak

\section{Introduction} 

In the past decade, social media have become one of the main sources of information for people around the world. Yet, using social media for news consumption is a double-edged sword. On the one hand, it offers low cost, easy access, and rapid dissemination. On the other hand, it comes with the danger of exposure to `fake news' containing poorly checked or even intentionally false information aimed at misleading and manipulating the readers to pursue certain political or economic agendas.   

The extensive spread of fake news has recently become a global problem and threat to modern democracies. The extensive spread of fake news before the United States 2016 presidential elections \cite{bovet2019influence} and the Brexit vote in United Kingdom has become the centerpiece of the controversy surrounding these political events and allegations of public opinion manipulation. Due the very high societal and economic cost of the phenomenon \cite{howell2013digital}, in the past year, fake news detection in social media has attracted enormous attention in the academic and industrial realms \cite{lazer2018science}.

Automatically detecting fake news poses challenges that defy existing content-based analysis approaches. One of the main reasons is that  often the interpretation of the news is highly nuanced and requires the knowledge of political or social context, or ``common sense'', which even the  currently most advanced natural language processing algorithms are still missing. 
Furthermore, fake news is often intentionally written by bad actors to appear as real news but containing false or manipulative information in ways that are hard even for trained human experts to detect.

\paragraph*{\bf Prior works. }
Existing approaches for fake news detection can be divided into three main categories, based on {\em content}, {\em social context}, and {\em propagation } \cite{shu2017fake,zhou2018fake}.
Content-based approaches, which are used in the majority of works on fake news detection, rely on linguistic (lexical and syntactical) features that can capture deceptive cues or writing styles \cite{afroz2012detecting,rubin2016fake,rashkin2017truth,potthast2017stylometric,perez2017automatic}. The main drawback of content-based approaches is that they can be defied by sufficiently sophisticated fake news that does not immediately appear as fake. Furthermore, most linguistic features are language-dependent, limiting the generality of these approaches.

Social context features include user demographics (such as age, gender, education, and political affiliation  \cite{shu2018understanding,long2017fake}), social network structure \cite{shu2019beyond,shu2019studying} (in the form of connections between users such as friendship or follower/followee relations) and user reactions (e.g. posts accompanying a news item \cite{ruchansky2017csi} or likes \cite{tacchini2017some}).

Propagation-based approaches are perhaps the most intriguing and promising research direction based on studying the news proliferation process over time. 
It has been argued that the fake news dissemination process is akin to infectious disease spread \cite{kucharski2016post} and can be understood with network epidemics models. 
There is substantial empirical evidence that fake news propagate differently from true news \cite{vosoughi2018spread} forming spreading patterns that could potentially be exploited for automatic fake news detection. 
%
%
%
By virtue of being content-agnostic, propagation-based features are likely generalizes across different languages, locales, and geographies, as opposed to content-based features that must be developed separately for each language. Furthermore, controlling the news spread patterns in a social network is generally beyond the capability of individual users, implying that propagation-based features would potentially be very hard to tamper with by adversarial attacks.


\paragraph*{\bf Main contribution.} 
So far, attempts to exploit news propagation for fake news detection applied `handcrafted' graph-theoretical features such as centrality, cliques, or connected components \cite{kwon2013prominent}. These features are rather arbitrary, too general, and not necessarily meaningful for the specific task of fake news detection. 
In this paper, we propose learning fake news specific propagation patterns by exploiting {\em geometric deep learning}, a novel class of deep learning methods designed to work on graph-structured data \cite{bronstein2017geometric}. 
Geometric deep learning naturally deals with heterogeneous data (such as user demographic and activity, social network structure, news propagation and content), thus carrying the potential of being a unifying framework for content, social context, and propagation based approaches. 
The model proposed in this paper is trained in a supervised manner on a large set of annotated fake and true stories spread on Twitter in the period 2013-2018. 
%
We perform extensive testing of our model in different challenging settings, showing that it achieves very high accuracy (nearly 93\% ROC AUC), requires very short news spread times (just a few hours of propagation), and performs well when the model is trained on data distant in time from the testing data. 
%



\section{Dataset}
\label{sec:dataset}

\begin{figure}[t]
    \centering
    \includegraphics[width=1.0\linewidth]{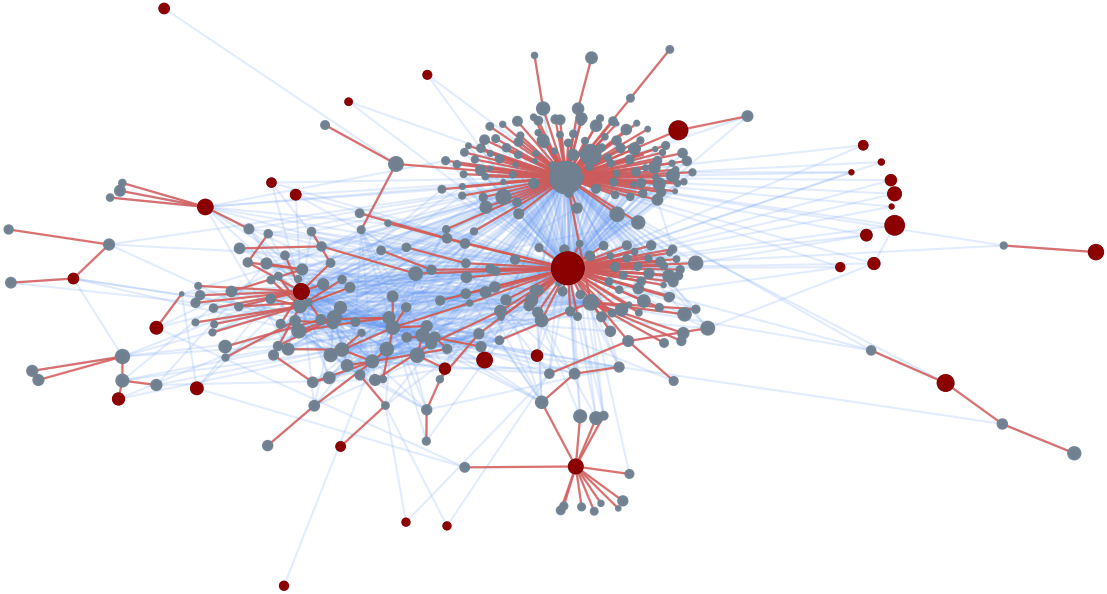}
    \caption{Example of a single news story spreading on a subset of the Twitter social network. Social connections between users are visualized as light-blue edges. A news URL is tweeted by multiple users (cascade roots denotes in red), each producing a cascade propagating over a subset of the social graph (red edges). Circle size represents the number of followers. Note that some cascades are small, containing only the root (the tweeting user) or just a few retweets. 
    %
    }
    \label{fig:cascade_spreadings}
\end{figure}

One of the key challenges in machine-learning based approaches in general, and in automatic fake news detection in particular, is collecting a sufficiently large, rich, and reliably labelled dataset on which the algorithms can be trained and tested.
Furthermore, the notion of `fake news' itself is rather vague and nuanced. 
To start with, there is no consensus as to what would be considered `news', let alone the label `true' or `false'. 
A large number of studies exploit the notion of reliable or unreliable {\em sources} as a proxy for true or false {\em stories}. While allowing to gather large datasets, such approaches have been criticized as too crude \cite{vosoughi2018spread}. 
In our study, we opted for a data collection process in which each `story' has an underlying article published on the web, and each such story is verified {\em individually}. 
In our classification of true or false statements we rely on professional non-profit journalist fact-checking organizations such as  Snopes,\footnote{\url{https://www.snopes.com/}}  PolitiFact,\footnote{\url{https://www.politifact.com/}} and Buzzfeed.\footnote{\url{https://www.buzzfeed.com/}} 
We note that our use of the term {\em fake news}, though disliked in the social science research community for its abuse in the political discourse, refers to both misinformation and disinformation, i.e. unintentional as well as deliberate spread of misleading or wrong narrative or facts.

\paragraph*{\bf Data collection protocol. }
Our data collection process was inspired by and largely followed the pioneering work of Vosoughi et al. \cite{vosoughi2018spread}. We used a collection of news verified by fact-checking organizations with established reputation in debunking rumors; 
each source fact-checking organization provides an archive of news with an associated short \emph{claim} (e.g. `Actress Allison Mack confessed that she sold children to the Rothschilds and Clintons') and a \emph{label} determining its veracity (`false' in the above example). First, we gathered the overall list of fact-checking articles  from such archives and, for simplicity, discarded claims with ambiguous labels, such as `mixed' or `partially true/false'.

Second, for each of the filtered articles we identified potentially related \emph{URLs} referenced by the fact-checkers, filtering out all those not mentioned at least once on Twitter. 
Third, trained human annotators were employed to ascertain whether the web pages associated with the collected URLs were \emph{matching} or \emph{denying} the claim, or were simply unrelated to that claim. 
This provided a simple method to propagate truth-labels from fact-checking verdicts to URLs: if a URL matches a claim, then it directly inherits the verdict; if it denies a claim, it inherits the opposite of the verdict (e.g. URLs matching a true claim are labeled as true, URLs denying a true claim are labeled as false). URLs gathered from different sources, with same veracity and date of first-appearance on Twitter were additionally inspected to ensure they referred to different articles. 

The last part of the data collection process consisted of the retrieval of Twitter data related to the propagation of news associated with a particular URL. Following the nomenclature of \cite{vosoughi2018spread}, we term as \emph{cascade} the news diffusion tree produced by a \emph{source} tweet referencing a URL and all of its retweets. 
For each URL, we searched for all the related cascades and enriched their Twitter-based characterization (i.e. users and tweet 
data) by drawing edges among users according to  Twitter's social network (see example in Figure~\ref{fig:cascade_spreadings}).

With regard to this last step of data collection, our approach is significantly different from the protocol of \cite{vosoughi2018spread}, where tweets linking to a fact-checking website were collected, thus essentially retrieving only cascades in which someone has posted a `proof-link' with the veracity of the news. 
Though significantly more laborious, we believe that our data collection protocol produces a much cleaner dataset.


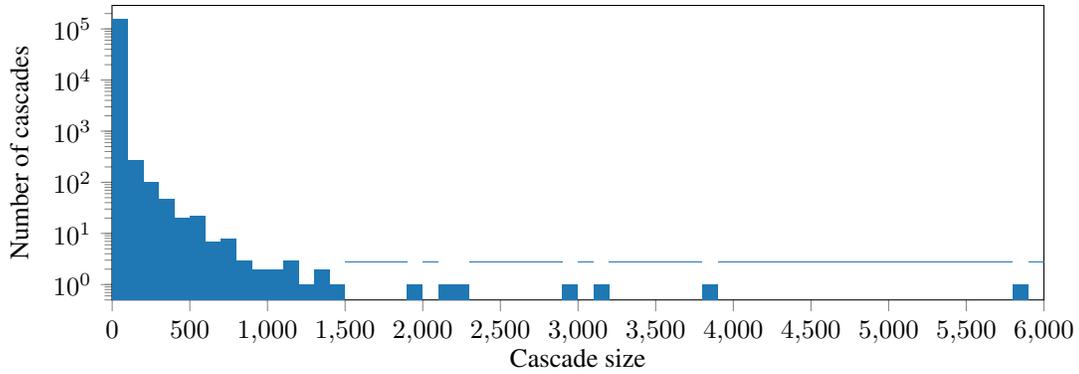
\begin{figure}[t]
    \centering
    \setlength\figureheight{5.5cm}
    \setlength\figurewidth{1.0\linewidth}
\begin{tikzpicture}

\definecolor{color0}{rgb}{0.12156862745098,0.466666666666667,0.705882352941177}

\begin{axis}[
height=\figureheight,
tick align=outside,
tick pos=left,
width=\figurewidth,
x grid style={lightgray!92.02614379084967!black},
xlabel={Cascade size},
xmin=-1, xmax=6000,
y grid style={lightgray!92.02614379084967!black},
ylabel={Number of cascades},
ymin=0.5, ymax=288326.113387547,
ymode=log
]
\draw[fill=color0,draw opacity=0] (axis cs:0,0.1) rectangle (axis cs:100,158449);
\draw[fill=color0,draw opacity=0] (axis cs:100,0.1) rectangle (axis cs:200,274);
\draw[fill=color0,draw opacity=0] (axis cs:200,0.1) rectangle (axis cs:300,102);
\draw[fill=color0,draw opacity=0] (axis cs:300,0.1) rectangle (axis cs:400,47);
\draw[fill=color0,draw opacity=0] (axis cs:400,0.1) rectangle (axis cs:500,20);
\draw[fill=color0,draw opacity=0] (axis cs:500,0.1) rectangle (axis cs:600,22);
\draw[fill=color0,draw opacity=0] (axis cs:600,0.1) rectangle (axis cs:700,7);
\draw[fill=color0,draw opacity=0] (axis cs:700,0.1) rectangle (axis cs:800,8);
\draw[fill=color0,draw opacity=0] (axis cs:800,0.1) rectangle (axis cs:900,3);
\draw[fill=color0,draw opacity=0] (axis cs:900,0.1) rectangle (axis cs:1000,2);
\draw[fill=color0,draw opacity=0] (axis cs:1000,0.1) rectangle (axis cs:1100,2);
\draw[fill=color0,draw opacity=0] (axis cs:1100,0.1) rectangle (axis cs:1200,3);
\draw[fill=color0,draw opacity=0] (axis cs:1200,0.1) rectangle (axis cs:1300,1);
\draw[fill=color0,draw opacity=0] (axis cs:1300,0.1) rectangle (axis cs:1400,2);
\draw[fill=color0,draw opacity=0] (axis cs:1400,0.1) rectangle (axis cs:1500,1);
\draw[fill=color0,draw opacity=0] (axis cs:1500,0) rectangle (axis cs:1600,0);
\draw[fill=color0,draw opacity=0] (axis cs:1600,0) rectangle (axis cs:1700,0);
\draw[fill=color0,draw opacity=0] (axis cs:1700,0) rectangle (axis cs:1800,0);
\draw[fill=color0,draw opacity=0] (axis cs:1800,0) rectangle (axis cs:1900,0);
\draw[fill=color0,draw opacity=0] (axis cs:1900,0.1) rectangle (axis cs:2000,1);
\draw[fill=color0,draw opacity=0] (axis cs:2000,0) rectangle (axis cs:2100,0);
\draw[fill=color0,draw opacity=0] (axis cs:2100,0.1) rectangle (axis cs:2200,1);
\draw[fill=color0,draw opacity=0] (axis cs:2200,0.1) rectangle (axis cs:2300,1);
\draw[fill=color0,draw opacity=0] (axis cs:2300,0) rectangle (axis cs:2400,0);
\draw[fill=color0,draw opacity=0] (axis cs:2400,0) rectangle (axis cs:2500,0);
\draw[fill=color0,draw opacity=0] (axis cs:2500,0) rectangle (axis cs:2600,0);
\draw[fill=color0,draw opacity=0] (axis cs:2600,0) rectangle (axis cs:2700,0);
\draw[fill=color0,draw opacity=0] (axis cs:2700,0) rectangle (axis cs:2800,0);
\draw[fill=color0,draw opacity=0] (axis cs:2800,0) rectangle (axis cs:2900,0);
\draw[fill=color0,draw opacity=0] (axis cs:2900,0.1) rectangle (axis cs:3000,1);
\draw[fill=color0,draw opacity=0] (axis cs:3000,0) rectangle (axis cs:3100,0);
\draw[fill=color0,draw opacity=0] (axis cs:3100,0.1) rectangle (axis cs:3200,1);
\draw[fill=color0,draw opacity=0] (axis cs:3200,0) rectangle (axis cs:3300,0);
\draw[fill=color0,draw opacity=0] (axis cs:3300,0) rectangle (axis cs:3400,0);
\draw[fill=color0,draw opacity=0] (axis cs:3400,0) rectangle (axis cs:3500,0);
\draw[fill=color0,draw opacity=0] (axis cs:3500,0) rectangle (axis cs:3600,0);
\draw[fill=color0,draw opacity=0] (axis cs:3600,0) rectangle (axis cs:3700,0);
\draw[fill=color0,draw opacity=0] (axis cs:3700,0) rectangle (axis cs:3800,0);
\draw[fill=color0,draw opacity=0] (axis cs:3800,0.1) rectangle (axis cs:3900,1);
\draw[fill=color0,draw opacity=0] (axis cs:3900,0) rectangle (axis cs:4000,0);
\draw[fill=color0,draw opacity=0] (axis cs:4000,0) rectangle (axis cs:4100,0);
\draw[fill=color0,draw opacity=0] (axis cs:4100,0) rectangle (axis cs:4200,0);
\draw[fill=color0,draw opacity=0] (axis cs:4200,0) rectangle (axis cs:4300,0);
\draw[fill=color0,draw opacity=0] (axis cs:4300,0) rectangle (axis cs:4400,0);
\draw[fill=color0,draw opacity=0] (axis cs:4400,0) rectangle (axis cs:4500,0);
\draw[fill=color0,draw opacity=0] (axis cs:4500,0) rectangle (axis cs:4600,0);
\draw[fill=color0,draw opacity=0] (axis cs:4600,0) rectangle (axis cs:4700,0);
\draw[fill=color0,draw opacity=0] (axis cs:4700,0) rectangle (axis cs:4800,0);
\draw[fill=color0,draw opacity=0] (axis cs:4800,0) rectangle (axis cs:4900,0);
\draw[fill=color0,draw opacity=0] (axis cs:4900,0) rectangle (axis cs:5000,0);
\draw[fill=color0,draw opacity=0] (axis cs:5000,0) rectangle (axis cs:5100,0);
\draw[fill=color0,draw opacity=0] (axis cs:5100,0) rectangle (axis cs:5200,0);
\draw[fill=color0,draw opacity=0] (axis cs:5200,0) rectangle (axis cs:5300,0);
\draw[fill=color0,draw opacity=0] (axis cs:5300,0) rectangle (axis cs:5400,0);
\draw[fill=color0,draw opacity=0] (axis cs:5400,0) rectangle (axis cs:5500,0);
\draw[fill=color0,draw opacity=0] (axis cs:5500,0) rectangle (axis cs:5600,0);
\draw[fill=color0,draw opacity=0] (axis cs:5600,0) rectangle (axis cs:5700,0);
\draw[fill=color0,draw opacity=0] (axis cs:5700,0) rectangle (axis cs:5800,0);
\draw[fill=color0,draw opacity=0] (axis cs:5800,0.1) rectangle (axis cs:5900,1);
\draw[fill=color0,draw opacity=0] (axis cs:5900,0) rectangle (axis cs:6000,0);
\draw[fill=color0,draw opacity=0] (axis cs:6000,0) rectangle (axis cs:6100,0);
\draw[fill=color0,draw opacity=0] (axis cs:6100,0) rectangle (axis cs:6200,0);
\draw[fill=color0,draw opacity=0] (axis cs:6200,0) rectangle (axis cs:6300,0);
\draw[fill=color0,draw opacity=0] (axis cs:6300,0) rectangle (axis cs:6400,0);
\end{axis}

\end{tikzpicture}
    \caption{Distribution of cascade sizes (number of tweets per cascade) in our dataset.
    }
    \label{fig:cascades}
\end{figure}


\begin{figure}[t]
    \setlength\figureheight{5.5cm}
    \setlength\figurewidth{1.0\linewidth}
    \begin{minipage}{1.0\linewidth}
    \centering
    \input{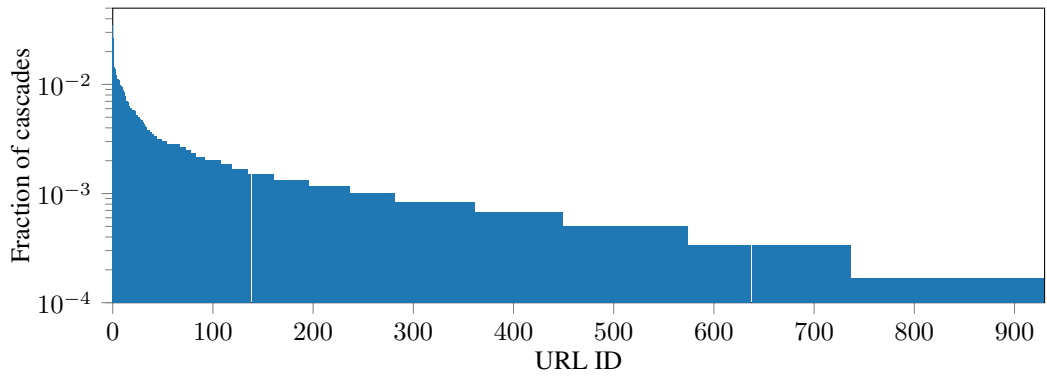}
    \end{minipage}
    \caption{Distribution of cascades over the 930 URLs available in our dataset with at least six tweets per cascade, sorted by the number cascades in descending order. The first 15 URLs (\char`\~1.5\% of the entire dataset)  correspond to 20\% of all the cascades.}
    \label{fig:URLs-distribution-cw-ds}
\end{figure}

\paragraph*{\bf Statistics.} 
Figures~\ref{fig:cascades}--\ref{fig:URLs-distribution-cw-ds} depict the statistics of our dataset. 
Overall, our collection consisted of $1,084$ labeled claims, spread on Twitter in $158,951$ cascades covering the period from May 2013 till January 2018. The total number of unique users involved in the spreading was $202,375$ and their respective social graph comprised $2,443,996$ edges. As we gathered $1,129$ URLs, the average number of article URLs per claim is around $1.04$; as such, a URL can be considered as a good proxy for a claim in our dataset and we will thus use the two terms synonymously hereinafter. We also note that, similarly to \cite{vosoughi2018spread}, a large proportion of cascades were of small size (the average number of tweets and users in a cascade is $2.79$, see also Figure~\ref{fig:cascades} depicting the distribution of cascade sizes), which required to use a threshold on a minimum cascade size for classifying these independently in some experiments (see details in Section \ref{sec:modperf}).
%



\paragraph*{\bf Features.} 
We extracted the following features describing news, users, and their activity, grouped into four categories: 
{\em User profile} (geolocalization and profile settings, language, word embedding of user profile self-description, date of account creation, and whether it has been verified), 
{\em User activity} (number of favorites, lists, and statuses), 
{\em Network and spreading} (social connections between the users, number of followers and friends, cascade spreading tree, retweet timestamps and source device, 
number of replies, quotes, favorites and retweets for the source tweet), and 
{\em Content} (word embedding of the tweet textual content and included hashtags).

\begin{figure}[t]
    \centering
    \includegraphics[width=1.0\linewidth]{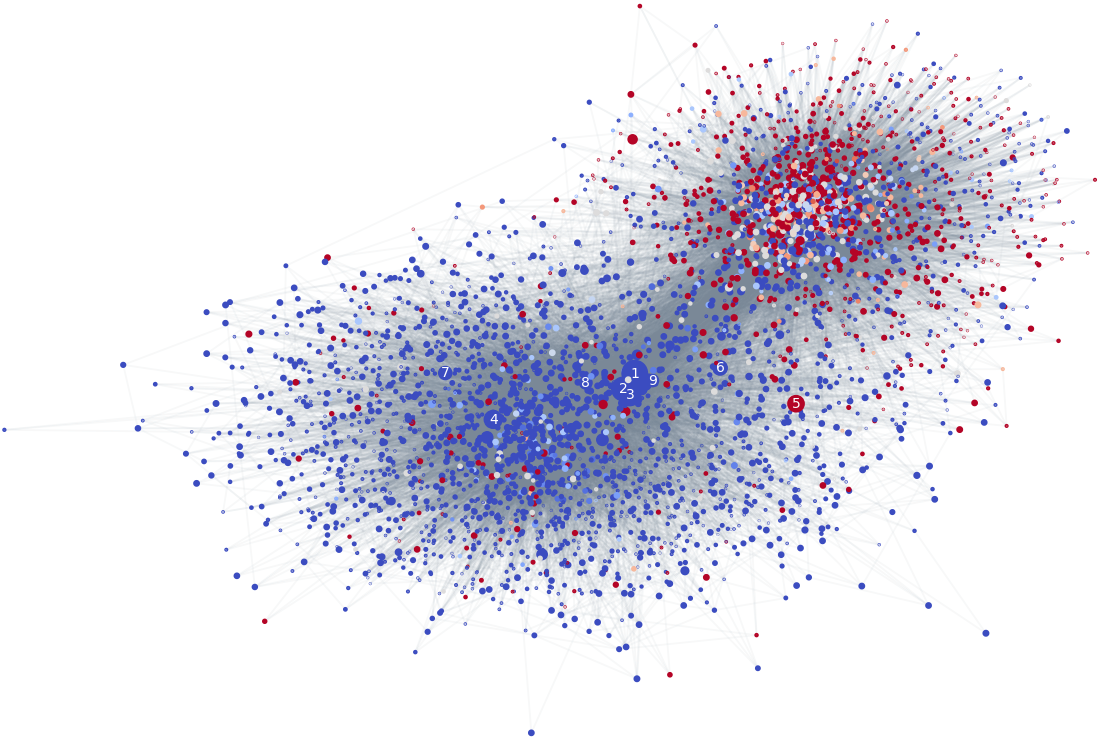}
    \caption{Subset of the Twitter network used in our study with estimated user credibility. Vertices represent users, gray edges the social connections. Vertex color and size encode the user credibility (blue = reliable, red = unreliable) and number of followers of each user, respectively. Numbers 1 to 9 represent the nine users with most followers.\vspace{-5mm}
    }
    \label{fig:polarization}
\end{figure}

\paragraph*{\bf Credibility and polarization. } The social network collected in our study manifests noticeable polarization depicted in Figure \ref{fig:polarization}. 
Each user in this plot is assigned a credibility score in the range $[-1, +1]$ computed as the difference between the proportion of (re)tweeted true and fake news (negative values representing fake are depicted in red; more credible users are represented in blue). 
%
%
The node positions of the graph are determined 
by topological embedding computed via the Fruchterman-Reingold force-directed algorithm~\cite{fruchterman1991graph}, grouping together nodes of the graph that are more strongly connected and mapping apart nodes that have weak connections.  
We observe that credible (blue) and non-credible (red) users tend to form two distinct communities, suggesting these two categories of tweeters prefer to have mostly homophilic interactions. While a deeper study of this phenomenon is beyond the scope of this paper, we note that similar polarization has been observed before in social networks, e.g. in the context of political discourse \cite{conover2011political} and might be related to `echo chamber' theories that attempt to explain the reasons for the difference in fake and true news propagation patterns. 

\section{Our model}
\label{sec:model}

\subsection{Geometric deep learning}

In the past decade, deep learning techniques have  had a remarkable impact on multiple domains, in particular computer vision, speech analysis, and natural language processing \cite{lecun2015deep}. 
However, most of popular deep neural models, such as convolutional neural networks (CNNs) \cite{lecun1998gradient}, are based on classical signal processing theory, with an underlying assumption of grid-structured (Euclidean) data. In recent years, there has been growing interest in generalizing deep learning techniques to non-Euclidean (graph- and manifold-structured) data. 
Early approaches to learning on graphs \cite{scarselli2009graph} predate the recent deep learning renaissance and are formulated as fixed points of learnable diffusion operators. 
The modern interest in deep learning on graphs can be attributed to the spectral CNN model of Bruna et al. \cite{bruna2014spectral}.  
Since some of the first works in this domain originated in graphics and geometry community  \cite{masci2015geodesic}, the term {\em geometric deep learning} is widely used as an umbrella term for non-Euclidean deep learning approaches \cite{bronstein2017geometric}.

Broadly speaking, graph CNNs replace the classical convolution operation on grids with a local permutation-invariant aggregation on the neighborhood of a vertex in a graph. In spectral graph CNNs \cite{bruna2014spectral}, this operation is performed in the spectral domain, by utilizing the analogy between the graph Laplacian eigenvectors and the classical Fourier transform; the filters are represented as learnable spectral coefficients. While conceptually important, spectral CNNs suffer from high computational complexity and difficulty to generalize across different domains \cite{bronstein2017geometric}. 
Follow-up works showed that the explicit eigendecomposition of the Laplacian can be avoided altogether by employing functions expressible in terms of simple matrix-vector operations, such as polynomials \cite{defferrard2016convolutional,kipf2016semi} or rational functions \cite{levie2017cayleynets}. 
Such spectral filters typically scale linearly with the graph size and can be generalized to higher order structures \cite{monti2018motifnet}, dual graphs (edge filters) \cite{monti2018dual}, and product graphs \cite{monti2017geometric}.

The Laplacian operator is only one example of fixed local permutation-invariant aggregation operation amounting to weighted averaging. More general operators have been proposed using edge convolutions \cite{wang2018dynamic}, neural message passing \cite{gilmer2017neural}, local charting \cite{monti2016geometric}, and graph attention \cite{velickovic2017graph}. 
On non-Euclidean domains with local low-dimensional structure (manifolds, meshes, point clouds), more powerful operators have been constructed using e.g. anisotropic diffusion kernels \cite{boscaini2016learning}. 


Being very abstract models of systems of relations and interactions, graphs naturally arise in various fields of science. For this reason, geometric deep learning techniques have been successfully applied across the board in problems such as 
computer graphics and vision \cite{boscaini2016learning,monti2016geometric,litany2018deformable,wang2018dynamic}, protection against adversarial attacks \cite{svoboda2018peernets}, recommendation systems \cite{monti2017geometric} quantum chemistry \cite{gilmer2017neural} and 
neutrino detection \cite{choma2018graph}, to mention a few.

%

\begin{figure}[t]
    \centering
    \includegraphics[width=1.0\linewidth]{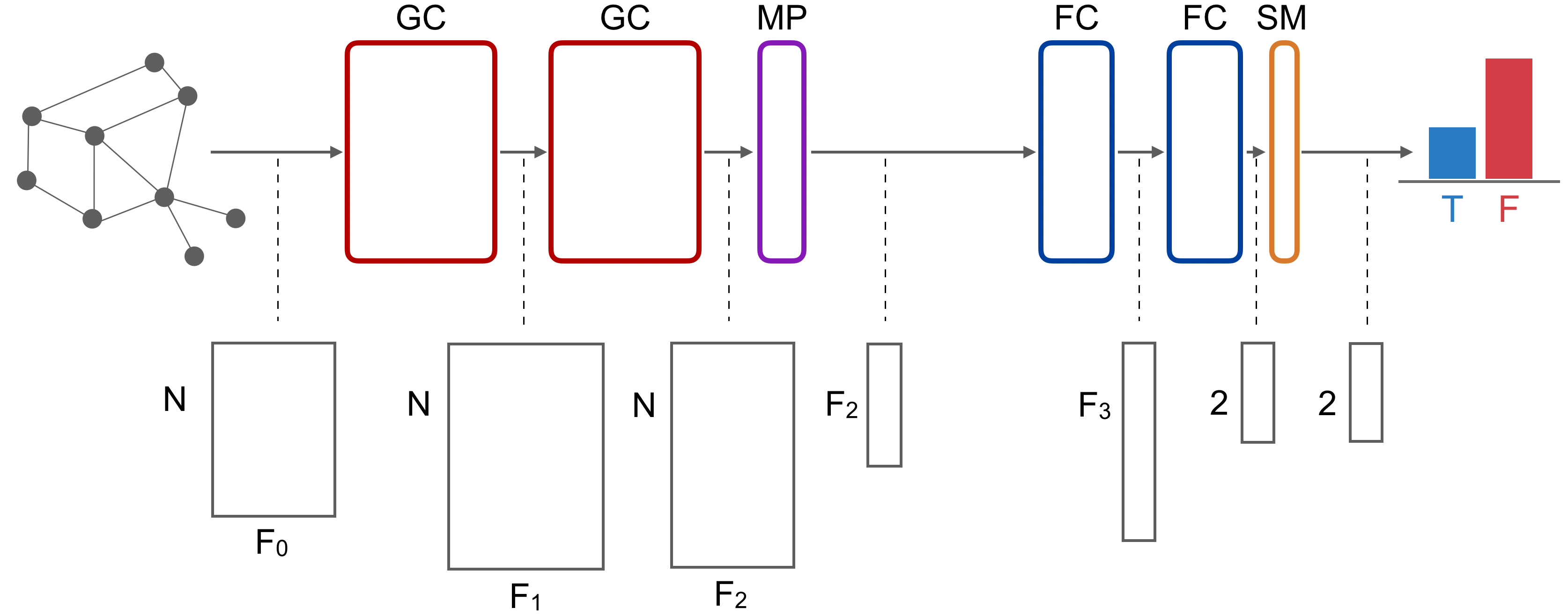}
    \caption{The architecture of our neural network model. Top row: GC = Graph Convolution, MP = Mean Pooling, FC = Fully Connected, SM = SoftMax layer. Bottom row: input/output tensors received/produced by each layer.
    }
    \label{fig:architecture}
\end{figure}

\subsection{Architecture and training settings}

Our deep learning model 
is described below. %
We used a four-layer Graph CNN with two convolutional layers (64-dimensional output features map in each) and two fully connected layers (producing 32- and 2-dimensional output features, respectively) to predict the fake/true class probabilities. Figure \ref{fig:architecture} depicts a block diagram of our model. 
One head of graph attention \cite{velickovic2017graph} was used in every convolutional layer to implement the filters together with mean-pooling for dimensionality reduction.
We used Scaled Exponential Linear Unit (SELU) \cite{klambauer2017self} as non-linearity throughout the entire network. Hinge loss was employed to train the neural network (we preferred hinge loss to the more commonly used mean cross entropy as it outperformed the latter in early experiments). No regularization was used with our model.  

\subsection{Input generation} 

Given a URL $u$ (or a cascade $c$ arising from $u$) with corresponding tweets $T_u = \{t_u^1, t_u^2,..., t_u^N\}$ mentioning it, we describe $u$ in terms of graph $G_u$. $G_u$ has tweets in $T_u$ as nodes and estimated news diffusion paths plus social relations as edges. 
In other words, given two nodes $i$ and $j$, edge $(i,j) \in G_u$ iff at least one of the following holds: $i$ \emph{follows} $j$ (i.e. the author of tweet $i$ follows the author of tweet $j$), $j$ \emph{follows} $i$, news spreading occurs from $i$ to $j$, or 
from $j$ to $i$.

News diffusion paths defining \emph{spreading trees} were estimated as in \cite{vosoughi2018spread} by jointly considering the timestamps of involved (re)tweets and the social connections between their authors. Given $t_u^n$ -- the retweet of a cascade related to URL $u$, and $\{t_u^0 \mathellipsis t_u^{n-1}\}$ -- the immediately preceding (re)tweets belonging to the same cascade and authored by users $\{a_u^0, \mathellipsis, a_u^{n}\}$, then:

\vspace{-2mm}
\begin{itemize}
    \item[1.] if $a_u^n$ \emph{follows} at least one user in $\{a_u^0, \mathellipsis, a_u^{n-1}\}$, we estimate news spreading to $t_u^n$ from the very last tweet in $\{t_u^0 \mathellipsis t_u^{n-1}\}$ whose author is \emph{followed} by ${a_u^n}$;
    \item[2.] if $a_u^n$ does not \emph{follow} any of the users in $\{a_u^0, \mathellipsis, a_u^{n-1}\}$, we conservatively estimate news spreading to $t_u^n$ from the user in $\{a_u^0, \mathellipsis, a_u^{n-1}\}$ having the largest number of followers (i.e. the most popular one).
\end{itemize}
\vspace{-2mm}

Finally, nodes and edges of graph $G_u$ have features describing them. Nodes, representing tweets and their authors, were characterized with all the features presented in Section \ref{sec:dataset}\footnote{For tweet content and user description embeddings we averaged together the embeddings of the constituent words (GloVe\cite{pennington2014glove} 200-dimensional vectors pre-trained on Twitter).}. As for edges, we used features representing the membership to each of the aforementioned four relations (\emph{following} and \emph{news spreading}, both directions). Our approach to defining graph connectivity and edge features allows, in graph convolution, to spread information independently of the relation direction while potentially giving different importance to the types of connections. Features of edge $(i,j)$ are concatenated to those of nodes $i$ and $j$ in the attention projection layer to achieve such behavior.

\section{Results}
\label{sec:results}

We considered two different settings of fake news detection: {\em URL-wise} and {\em cascade-wise}, using the same architecture for both settings. In the first setting, we attempted to predict the true/fake label of a URL containing a news story from all the Twitter cascades it generated. On average, each URL resulted in approximately 141 cascades. 
In the latter setting, which is significantly more challenging, we assumed to be given only one cascade arising from a URL and attempted to predict the  label associated with that URL. Our assumption is that all the cascades associated with a URL inherit the label of the latter. While we checked this assumption to be true in most cases in our dataset, it is possible that an article is for example tweeted with a comment denying its content. 
We leave the analysis of comments accompanying tweets/retweets as a future research direction.

\subsection{Model performance}
\label{sec:modperf}

For URL-wise classification, we used five randomized training/test/validation splits.  
On average, the training, test, and validation sets contained 677, 226, and 226 URLs, respectively, with 83.26\% true and 16.74\% false labels ($\pm$ 0.06\% and 0.15\% for training and validation/test set respectively). 
%
%
%
For cascade-wise classification 
we used the same split initially realized for URL-wise classification (i.e. all cascades originated by URL $u$ are placed in the same fold as $u$). 
Cascades containing less than 6 tweets were discarded; the reason for the choice of this threshold is motivated below. 
Full cascade duration (24 hr) was used for both settings of this experiment. 
The training, test, and validation sets contained on average 3586, 1195, 1195 cascades, respectively, with 81.73\% true and 18.27\% false labels ($\pm$ 3.25\% and 6.50\% for training and validation/test set respectively). 
%
%

Our neural network was trained for $25\times 10^3$ and $50\times 10^3$ iterations in the URL- and cascade-wise settings, respectively, using AMSGrad \cite{reddi2018convergence} with learning rate of $5 \times 10^{-4}$ and mini-batch of size one.

\begin{figure}[t]
    \centering
    \setlength\figureheight{6cm}
    \setlength\figurewidth{1.0\linewidth}
    \input{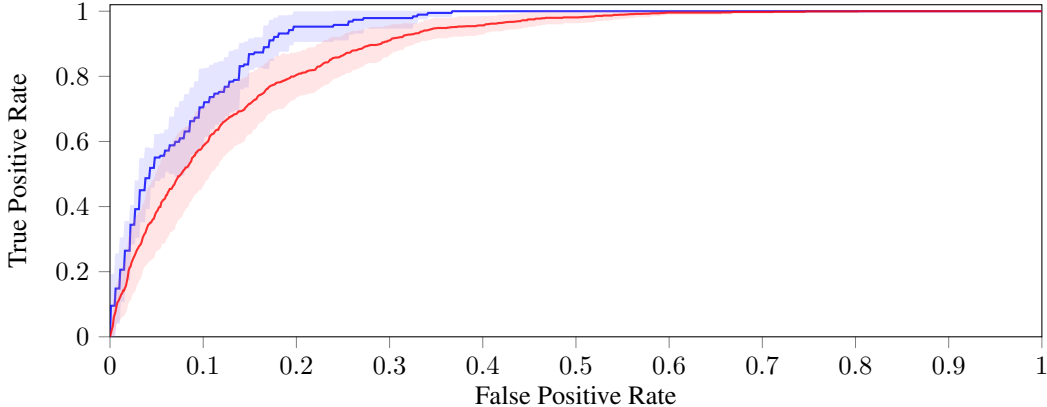}
    \caption{Performance of URL-wise (blue) and cascade-wise (red) fake news detection using 24hr-long diffusion time. Shown are ROC curves averaged on five folds (the shaded areas represent the standard deviations). ROC AUC is 92.70 $\pm$ 1.80\% for URL-wise classification and 88.30 $\pm$ 2.74\% for cascade-wise classification, respectively. Only cascades with at least 6 tweets were considered for cascade-wise classification.}
    \label{fig:roc_24h_5_folds}
\end{figure}

Figure~\ref{fig:roc_24h_5_folds} depicts the performance of URL- (blue) and cascade-wise (red) fake news classification represented as a tradeoff (ROC curve) between false positive rate (fraction of true news wrongly classified as fake) and true positive rate (fraction of fake news correctly classified as fake). We use {\em area under the ROC curve} (ROC AUC) as an aggregate measure of accuracy. On the above splits, our method achieved mean ROC AUC of $92.70 \pm 1.80\%$ and $88.30 \pm 2.74\%$ in the URL- and cascade-wise settings, respectively. 

\begin{figure}[t]
    \centering
    \includegraphics[width=1.0\linewidth]{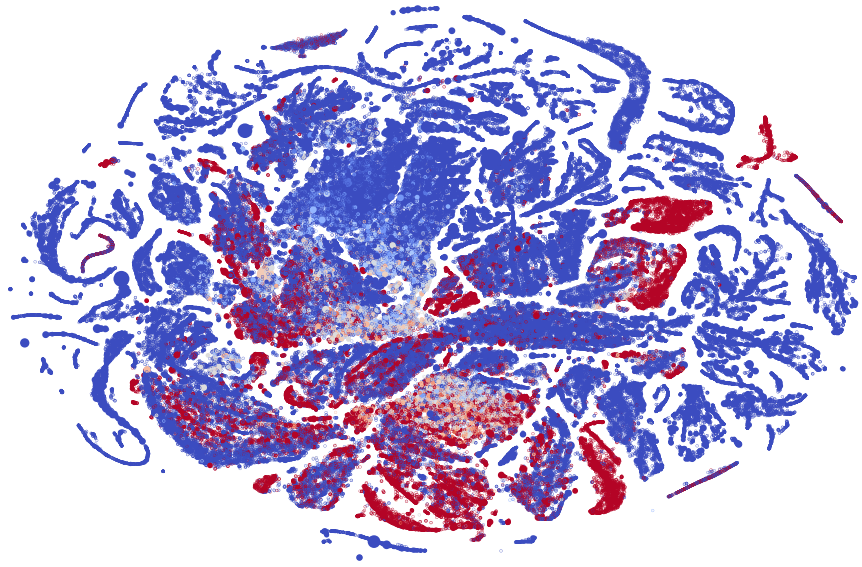}

    \caption{T-SNE  
    embedding of the vertex-wise features produced by our neural network at the last convolutional layer representing all the users in our study, color-coded according to their credibility (blue = reliable, red = unreliable). Clusters of users with different credibility clearly emerge, indicative that the neural network learns features useful for fake news detection.}
    \label{fig:t-sne-embedding}
\end{figure}

Figure~\ref{fig:t-sne-embedding} depicts a low-dimensional plot of the last graph convolutional layer vertex-wise features obtained using t-SNE embedding. The vertices are colored using the credibility score defined in Section~\ref{sec:dataset}. We observe clear clusters of reliable (blue) and unreliable (red) users, which is indicative of the neural network learning features that are useful for fake news classification.

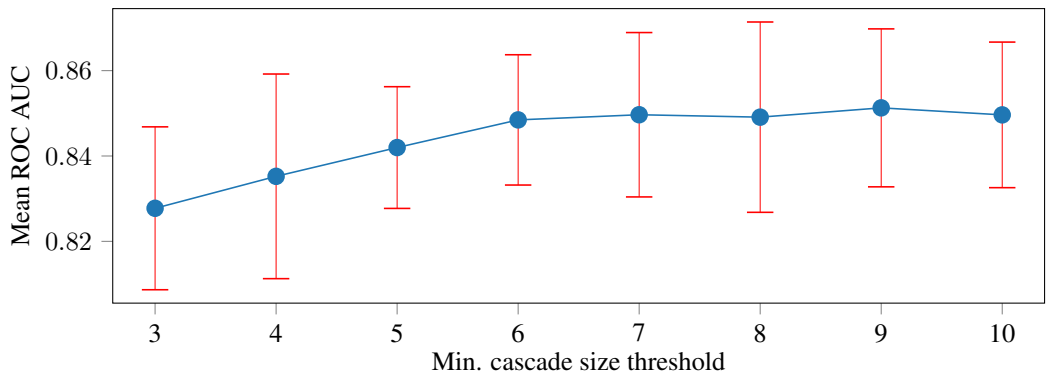
\begin{figure}[t]
    \setlength\figureheight{5.5cm}
    \setlength\figurewidth{1.0\linewidth}
    \begin{minipage}{1.0\linewidth}
    \centering
\begin{tikzpicture}

\definecolor{color0}{rgb}{0.12156862745098,0.466666666666667,0.705882352941177}

\begin{axis}[
height=\figureheight,
tick align=outside,
tick pos=left,
width=\figurewidth,
x grid style={white!69.01960784313725!black},
xlabel={Min. cascade size threshold},
xmin=-0.35, xmax=7.35,
xtick={0,1,2,3,4,5,6,7},
xticklabels={3,4,5,6,7,8,9,10},
y grid style={white!69.01960784313725!black},
ylabel={Mean ROC AUC},
ymin=0.805521523124913, ymax=0.874536993853357
]
\path [draw=red, very thin] (axis cs:0,0.808658589976206)
--(axis cs:0,0.846829862519869);

\path [draw=red, very thin] (axis cs:1,0.811255408959556)
--(axis cs:1,0.859209422269303);

\path [draw=red, very thin] (axis cs:2,0.827716592331318)
--(axis cs:2,0.856237633109426);

\path [draw=red, very thin] (axis cs:3,0.833193759499748)
--(axis cs:3,0.863724919470207);

\path [draw=red, very thin] (axis cs:4,0.830407948745744)
--(axis cs:4,0.868925303330902);

\path [draw=red, very thin] (axis cs:5,0.826800452861971)
--(axis cs:5,0.871399927002064);

\path [draw=red, very thin] (axis cs:6,0.832772465038802)
--(axis cs:6,0.869786530565175);

\path [draw=red, very thin] (axis cs:7,0.832577092716075)
--(axis cs:7,0.86669237558655);

\addplot [semithick, red, mark=-, mark size=5, mark options={solid}, only marks, forget plot]
table [row sep=\\]{%
0	0.808658589976206 \\
1	0.811255408959556 \\
2	0.827716592331318 \\
3	0.833193759499748 \\
4	0.830407948745744 \\
5	0.826800452861971 \\
6	0.832772465038802 \\
7	0.832577092716075 \\
};
\addplot [semithick, red, mark=-, mark size=5, mark options={solid}, only marks, forget plot]
table [row sep=\\]{%
0	0.846829862519869 \\
1	0.859209422269303 \\
2	0.856237633109426 \\
3	0.863724919470207 \\
4	0.868925303330902 \\
5	0.871399927002064 \\
6	0.869786530565175 \\
7	0.86669237558655 \\
};
\addplot [semithick, color0, solid, mark=*, mark size=3, mark options={solid}, forget plot]
table [row sep=\\]{%
0	0.827744226248037 \\
1	0.835232415614429 \\
2	0.841977112720372 \\
3	0.848459339484977 \\
4	0.849666626038323 \\
5	0.849100189932018 \\
6	0.851279497801989 \\
7	0.849634734151313 \\
};
\end{axis}

\end{tikzpicture}
    \end{minipage}
    \caption{Performance of cascade-wise fake news detection (mean ROC AUC, averaged on five folds) using minimum cascade size threshold. Best performance is obtained by filtering out cascades smaller than 6 tweets. }
    \label{fig:cw-min-thresh}
\end{figure}

{\bf Influence of minimum cascade size.} 
One of the characteristics of our dataset (as well as the dataset in the study of \cite{vosoughi2018spread}) is the abundance of small cascades containing just a few users (see Figure~\ref{fig:cascades}). Since our approach relies on the spreading of news across the Twitter social network, such examples may be hard to classify, as too small cascades may manifest no clear diffusion pattern. 
To identify the minimum useful cascade size, we investigated the performance of our model in the cascade-wise classification setting using cascades of various minimum sizes (Figure \ref{fig:cw-min-thresh}).
%
%
As expected, the model performance increases with larger cascades, reaching saturation for cascades of at least 6 tweets (leaving a total of 5,976 samples). 
This experiment motivates our choice of using 6 tweets as the minimum cascade size in cascade-wise experiments in our study.

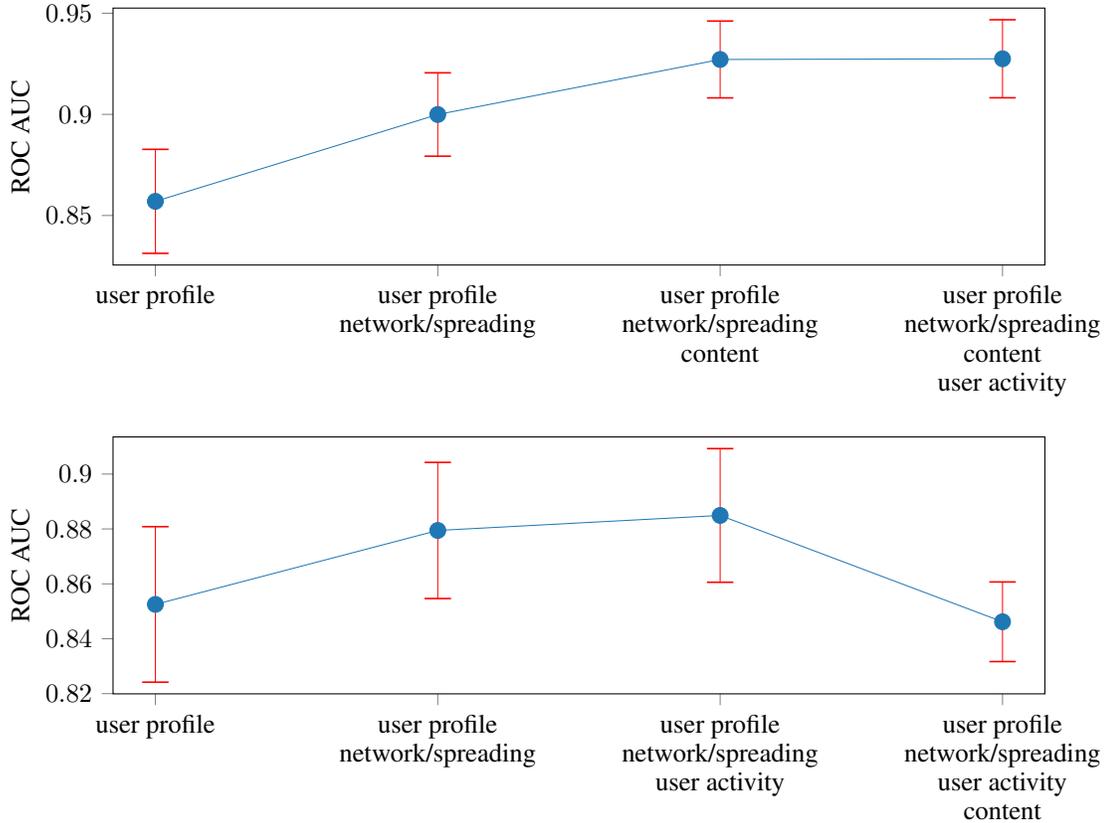
\begin{figure}[t]
    \centering
    \setlength\figureheight{5cm}
    \setlength\figurewidth{1.0\linewidth}
    \begin{minipage}{1.0\linewidth}
\begin{tikzpicture}

\definecolor{color0}{rgb}{0.12156862745098,0.466666666666667,0.705882352941177}

\begin{axis}[
height=\figureheight,
tick align=outside,
tick pos=left,
width=\figurewidth,
x grid style={white!69.01960784313725!black},
xmin=-0.15, xmax=3.15,
xtick={0,1,2,3},
xticklabel style={align=center},
xticklabels={
    user profile,
    user profile \\ network/spreading,
    user profile \\ network/spreading \\ content,
    user profile \\ network/spreading \\ content \\ user activity },
y grid style={white!69.01960784313725!black},
ylabel={ROC AUC},
ymin=0.825520280143442, ymax=0.952540216733725
]
\path [draw=red, very thin] (axis cs:0,0.831293913624818)
--(axis cs:0,0.882702333462134);

\path [draw=red, very thin] (axis cs:1,0.879328309459673)
--(axis cs:1,0.920532469572438);

\path [draw=red, very thin] (axis cs:2,0.908141505303767)
--(axis cs:2,0.94612289347351);

\path [draw=red, very thin] (axis cs:3,0.908193920364371)
--(axis cs:3,0.946766583252348);

\addplot [semithick, red, mark=-, mark size=5, mark options={solid}, only marks, forget plot]
table [row sep=\\]{%
0	0.831293913624818 \\
1	0.879328309459673 \\
2	0.908141505303767 \\
3	0.908193920364371 \\
};
\addplot [semithick, red, mark=-, mark size=5, mark options={solid}, only marks, forget plot]
table [row sep=\\]{%
0	0.882702333462134 \\
1	0.920532469572438 \\
2	0.94612289347351 \\
3	0.946766583252348 \\
};
\addplot [line width=0.38pt, color0, solid, mark=*, mark size=3, mark options={solid}, forget plot]
table [row sep=\\]{%
0	0.856998123543476 \\
1	0.899930389516056 \\
2	0.927132199388638 \\
3	0.927480251808359 \\
};
\end{axis}

\end{tikzpicture}
    \end{minipage} \vspace{2mm}
    \hfill
    \begin{minipage}{1.0\linewidth}
    \setlength\figurewidth{1.0\linewidth}
\begin{tikzpicture}

\definecolor{color0}{rgb}{0.12156862745098,0.466666666666667,0.705882352941177}

\begin{axis}[
height=\figureheight,
tick align=outside,
tick pos=left,
width=\figurewidth,
x grid style={white!69.01960784313725!black},
xmin=-0.15, xmax=3.15,
xtick={0,1,2,3},
xticklabel style={align=center},
xticklabels={
    user profile,
    user profile \\ network/spreading,
    user profile \\ network/spreading \\ user activity,
    user profile \\ network/spreading \\ user activity \\ content},
y grid style={white!69.01960784313725!black},
ylabel={ROC AUC},
ymin=0.819925865918205, ymax=0.913523598628579
]
\path [draw=red, very thin] (axis cs:0,0.824180308314131)
--(axis cs:0,0.880831097841261);

\path [draw=red, very thin] (axis cs:1,0.854648818753766)
--(axis cs:1,0.904254000153707);

\path [draw=red, very thin] (axis cs:2,0.860576119454199)
--(axis cs:2,0.909269156232653);

\path [draw=red, very thin] (axis cs:3,0.831677941311231)
--(axis cs:3,0.860688746994686);

\addplot [semithick, red, mark=-, mark size=5, mark options={solid}, only marks, forget plot]
table [row sep=\\]{%
0	0.824180308314131 \\
1	0.854648818753766 \\
2	0.860576119454199 \\
3	0.831677941311231 \\
};
\addplot [semithick, red, mark=-, mark size=5, mark options={solid}, only marks, forget plot]
table [row sep=\\]{%
0	0.880831097841261 \\
1	0.904254000153707 \\
2	0.909269156232653 \\
3	0.860688746994686 \\
};
\addplot [line width=0.38pt, color0, solid, mark=*, mark size=3, mark options={solid}, forget plot]
table [row sep=\\]{%
0	0.852505703077696 \\
1	0.879451409453736 \\
2	0.884922637843426 \\
3	0.846183344152958 \\
};
\end{axis}

\end{tikzpicture}
    \end{minipage}
    \caption{Ablation study result on URL-wise (top) / cascade-wise (bottom) fake news detection, using backward feature selection. Shown is performance (ROC AUC) for 
    our model trained on subsets of features, grouped into four categories: user profile, network and spreading, content, and user activity. Groups are sorted for importance from left to right.
    }
    \label{fig:ablation_study}
\end{figure}

{\bf Ablation study.} To further highlight the importance of the different categories of features provided as input to the model, we conducted an ablation study by means of backward-feature selection. 
We considered four groups of features defined in Section~\ref{sec:dataset}:  {\em user profile}, 
{\em user activity}, 
{\em network and spreading}, and 
{\em content}.  
The results of ablation experiment are shown in 
Figure \ref{fig:ablation_study} for the URL- (top) and cascade-wise (bottom) settings. 
In both settings, user-profile and network/spreading appear as the two most important feature groups, and allow achieving satisfactory classification results (near $90\%$ ROC AUC) with the proposed model. 


Interestingly, in the cascade-wise setting, while all features were positively contributing to the final predictions at URL-level, removing tweet content from the provided input improves performance by $4\%$. 
This seemingly contradictory result can be explained by looking at the distribution of cascades over all the available URLs (Figure \ref{fig:URLs-distribution-cw-ds}): $20\%$ of cascades are associated to the top 15 largest URLs in our dataset ($\sim 1.5\%$ out of a total of 930). Since tweets citing the same URL typically present similar content, it is easy for the model to overfit on this particular feature. Proper regularization (e.g. dropout or weight decay) should thus be introduced to avoid overfitting and improve performance at test time. 
We leave this further study for future research. 
For simplicity, by leveraging the capabilities of our model to classify fake news in a content-free scenario, we decided to completely ignore content-based descriptors (tweet word embeddings) for cascade-wise classification and let the model exploit only user- and propagation-related features. 


\subsection{News spreading over time}

One of the key differentiators of propagation-based methods from their content-based counterparts, namely relying on the news spreading features,  potentially raises the following question: for how much time do the news have to spread before we can classify them reliably?  
We conducted a series of experiments to study the extent to which this is the case with our approach.

For this purpose, we truncated the cascades after time $t$ starting from the first tweet, with $t$ varying from 0 (effectively considering only the initial tweet, i.e. the `root' of each cascade) to 24 hours (the full cascade duration) with one hour increments. 
The model was trained separately for each value of $t$. 
Five-fold cross validation was used to reduce the bias of the estimations while containing the overall computational cost.

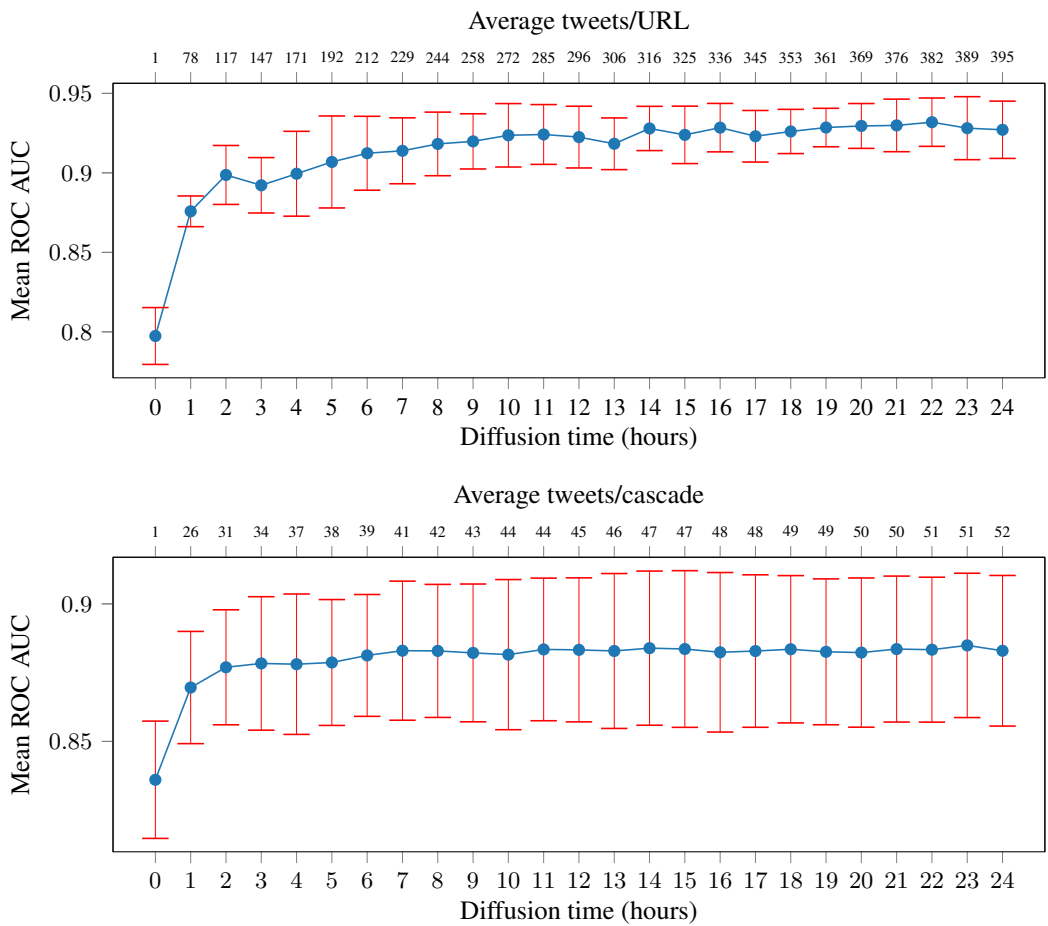
\begin{figure}[t]
    \centering
    
    \setlength\figureheight{5.5cm}
    \setlength\figurewidth{1.0\linewidth}
    \begin{minipage}{1.0\linewidth}
\begin{tikzpicture}

\definecolor{color0}{rgb}{0.12156862745098,0.466666666666667,0.705882352941177}

\begin{axis}[
height=\figureheight,
tick align=outside,
tick pos=left,
width=\figurewidth,
x grid style={lightgray!92.02614379084967!black},
xlabel={Diffusion time (hours)},
xmin=-0.2, xmax=26.2,
xtick={1,2,3,4,5,6,7,8,9,10,11,12,13,14,15,16,17,18,19,20,21,22,23,24,25},
xticklabels={0,1,2,3,4,5,6,7,8,9,10,11,12,13,14,15,16,17,18,19,20,21,22,23,24},
ticklabel style = {font=\small},
y grid style={lightgray!92.02614379084967!black},
ylabel={Mean ROC AUC},
ymin=0.771114826322943, ymax=0.956327738648662
]
\path [draw=red, very thin] (axis cs:1,0.779533595065022)
--(axis cs:1,0.815286476966697);

\path [draw=red, very thin] (axis cs:2,0.866160491740107)
--(axis cs:2,0.885482920989532);

\path [draw=red, very thin] (axis cs:3,0.880148638627168)
--(axis cs:3,0.917219782425463);

\path [draw=red, very thin] (axis cs:4,0.874768480406611)
--(axis cs:4,0.909602452676528);

\path [draw=red, very thin] (axis cs:5,0.87276907429466)
--(axis cs:5,0.926112618148062);

\path [draw=red, very thin] (axis cs:6,0.877934635533541)
--(axis cs:6,0.935786196160415);

\path [draw=red, very thin] (axis cs:7,0.889103666462403)
--(axis cs:7,0.935567197010252);

\path [draw=red, very thin] (axis cs:8,0.893151885286405)
--(axis cs:8,0.934590919108136);

\path [draw=red, very thin] (axis cs:9,0.898209015979334)
--(axis cs:9,0.938197267123477);

\path [draw=red, very thin] (axis cs:10,0.902441635851724)
--(axis cs:10,0.937162189698351);

\path [draw=red, very thin] (axis cs:11,0.90367939417884)
--(axis cs:11,0.943569478433974);

\path [draw=red, very thin] (axis cs:12,0.90533095337467)
--(axis cs:12,0.942948457054796);

\path [draw=red, very thin] (axis cs:13,0.903086080536284)
--(axis cs:13,0.941903477891124);

\path [draw=red, very thin] (axis cs:14,0.902036245348434)
--(axis cs:14,0.934542550692848);

\path [draw=red, very thin] (axis cs:15,0.914015198531541)
--(axis cs:15,0.941841161748112);

\path [draw=red, very thin] (axis cs:16,0.905839525635858)
--(axis cs:16,0.941948071591829);

\path [draw=red, very thin] (axis cs:17,0.91320815473994)
--(axis cs:17,0.943666637185244);

\path [draw=red, very thin] (axis cs:18,0.906796071668923)
--(axis cs:18,0.939234617474869);

\path [draw=red, very thin] (axis cs:19,0.912130902467997)
--(axis cs:19,0.939933199708088);

\path [draw=red, very thin] (axis cs:20,0.916412973857931)
--(axis cs:20,0.940596499220971);

\path [draw=red, very thin] (axis cs:21,0.91539886917699)
--(axis cs:21,0.943597529237101);

\path [draw=red, very thin] (axis cs:22,0.913310757835243)
--(axis cs:22,0.946366612704389);

\path [draw=red, very thin] (axis cs:23,0.91667587146399)
--(axis cs:23,0.947073712386378);

\path [draw=red, very thin] (axis cs:24,0.908295442792789)
--(axis cs:24,0.947908969906584);

\path [draw=red, very thin] (axis cs:25,0.909086933151419)
--(axis cs:25,0.945103315327743);

\addplot [semithick, red, mark=-, mark size=5, mark options={solid}, only marks, forget plot]
table [row sep=\\]{%
1	0.779533595065022 \\
2	0.866160491740107 \\
3	0.880148638627168 \\
4	0.874768480406611 \\
5	0.87276907429466 \\
6	0.877934635533541 \\
7	0.889103666462403 \\
8	0.893151885286405 \\
9	0.898209015979334 \\
10	0.902441635851724 \\
11	0.90367939417884 \\
12	0.90533095337467 \\
13	0.903086080536284 \\
14	0.902036245348434 \\
15	0.914015198531541 \\
16	0.905839525635858 \\
17	0.91320815473994 \\
18	0.906796071668923 \\
19	0.912130902467997 \\
20	0.916412973857931 \\
21	0.91539886917699 \\
22	0.913310757835243 \\
23	0.91667587146399 \\
24	0.908295442792789 \\
25	0.909086933151419 \\
};
\addplot [semithick, red, mark=-, mark size=5, mark options={solid}, only marks, forget plot]
table [row sep=\\]{%
1	0.815286476966697 \\
2	0.885482920989532 \\
3	0.917219782425463 \\
4	0.909602452676528 \\
5	0.926112618148062 \\
6	0.935786196160415 \\
7	0.935567197010252 \\
8	0.934590919108136 \\
9	0.938197267123477 \\
10	0.937162189698351 \\
11	0.943569478433974 \\
12	0.942948457054796 \\
13	0.941903477891124 \\
14	0.934542550692848 \\
15	0.941841161748112 \\
16	0.941948071591829 \\
17	0.943666637185244 \\
18	0.939234617474869 \\
19	0.939933199708088 \\
20	0.940596499220971 \\
21	0.943597529237101 \\
22	0.946366612704389 \\
23	0.947073712386378 \\
24	0.947908969906584 \\
25	0.945103315327743 \\
};
\addplot [semithick, color0, mark=*, mark size=2, solid, mark options={solid}, forget plot]
table [row sep=\\]{%
1	0.797410036015859 \\
2	0.875821706364819 \\
3	0.898684210526316 \\
4	0.89218546654157 \\
5	0.899440846221361 \\
6	0.906860415846978 \\
7	0.912335431736328 \\
8	0.91387140219727 \\
9	0.918203141551406 \\
10	0.919801912775037 \\
11	0.923624436306407 \\
12	0.924139705214733 \\
13	0.922494779213704 \\
14	0.918289398020641 \\
15	0.927928180139826 \\
16	0.923893798613843 \\
17	0.928437395962592 \\
18	0.923015344571896 \\
19	0.926032051088042 \\
20	0.928504736539451 \\
21	0.929498199207046 \\
22	0.929838685269816 \\
23	0.931874791925184 \\
24	0.928102206349687 \\
25	0.927095124239581 \\
};
\end{axis}

\begin{axis}[
axis x line=top,
x axis line style=-,
height=\figureheight,
tick align=outside,
width=\figurewidth,
x grid style={lightgray!92.02614379084967!black},
xlabel={Average tweets/URL},
xmin=-0.2, xmax=26.2,
xtick pos=right,
xtick={1,2,3,4,5,6,7,8,9,10,11,12,13,14,15,16,17,18,19,20,21,22,23,24,25},
xticklabels={1,78,117,147,171,192,212,229,244,258,272,285,296,306,316,325,336,345,353,361,369,376,382,389,395},
ticklabel style = {font=\tiny},
y grid style={lightgray!92.02614379084967!black},
yticklabel=\empty,
ymin=0.771114826322943, ymax=0.956327738648662,
ytick pos=left
]
\end{axis}

\end{tikzpicture}
    \end{minipage}\vspace{2mm}
    \begin{minipage}{1.0\linewidth}
\begin{tikzpicture}

\definecolor{color0}{rgb}{0.12156862745098,0.466666666666667,0.705882352941177}

\begin{axis}[
height=\figureheight,
tick align=outside,
tick pos=left,
width=\figurewidth,
xtick={0,1,2,3,4,5,6,7,8,9,10,11,12,13,14,15,16,17,18,19,20,21,22,23,24},
x grid style={white!69.01960784313725!black},
ticklabel style = {font=\small},
xlabel={Diffusion time (hours)},
xmin=-1.2, xmax=25.2,
y grid style={white!69.01960784313725!black},
ylabel={Mean ROC AUC},
ymin=0.809849383407478, ymax=0.916956220480781
]
\path [draw=red, very thin] (axis cs:0,0.814717876001719)
--(axis cs:0,0.857370988570784);

\path [draw=red, very thin] (axis cs:1,0.849181700890661)
--(axis cs:1,0.889989837071926);

\path [draw=red, very thin] (axis cs:2,0.85603398477765)
--(axis cs:2,0.897857536515589);

\path [draw=red, very thin] (axis cs:3,0.854082544834216)
--(axis cs:3,0.902615146422788);

\path [draw=red, very thin] (axis cs:4,0.852538904466708)
--(axis cs:4,0.903599049030277);

\path [draw=red, very thin] (axis cs:5,0.855803540008657)
--(axis cs:5,0.901577060806772);

\path [draw=red, very thin] (axis cs:6,0.859103992708925)
--(axis cs:6,0.903414352156707);

\path [draw=red, very thin] (axis cs:7,0.857696149577344)
--(axis cs:7,0.908273244201654);

\path [draw=red, very thin] (axis cs:8,0.858712222391874)
--(axis cs:8,0.907093797343241);

\path [draw=red, very thin] (axis cs:9,0.85712726324313)
--(axis cs:9,0.907228283080604);

\path [draw=red, very thin] (axis cs:10,0.854244961150167)
--(axis cs:10,0.908851044286626);

\path [draw=red, very thin] (axis cs:11,0.857508817220528)
--(axis cs:11,0.90937092351042);

\path [draw=red, very thin] (axis cs:12,0.857094053154282)
--(axis cs:12,0.909473433566874);

\path [draw=red, very thin] (axis cs:13,0.854709907843642)
--(axis cs:13,0.911041896287881);

\path [draw=red, very thin] (axis cs:14,0.855874121881354)
--(axis cs:14,0.911937136210155);

\path [draw=red, very thin] (axis cs:15,0.85510786797146)
--(axis cs:15,0.91208772788654);

\path [draw=red, very thin] (axis cs:16,0.853394178417712)
--(axis cs:16,0.911409317154316);

\path [draw=red, very thin] (axis cs:17,0.855131034561963)
--(axis cs:17,0.910581078374351);

\path [draw=red, very thin] (axis cs:18,0.856725221169669)
--(axis cs:18,0.910290850323322);

\path [draw=red, very thin] (axis cs:19,0.856050802350276)
--(axis cs:19,0.909101889906635);

\path [draw=red, very thin] (axis cs:20,0.855161766759616)
--(axis cs:20,0.909430289783604);

\path [draw=red, very thin] (axis cs:21,0.857026929938738)
--(axis cs:21,0.910113561736539);

\path [draw=red, very thin] (axis cs:22,0.857003395942089)
--(axis cs:22,0.909703632498501);

\path [draw=red, very thin] (axis cs:23,0.858666350430549)
--(axis cs:23,0.911146831777076);

\path [draw=red, very thin] (axis cs:24,0.855571254263224)
--(axis cs:24,0.910342849895613);

\addplot [semithick, red, mark=-, mark size=5, mark options={solid}, only marks, forget plot]
table [row sep=\\]{%
0	0.814717876001719 \\
1	0.849181700890661 \\
2	0.85603398477765 \\
3	0.854082544834216 \\
4	0.852538904466708 \\
5	0.855803540008657 \\
6	0.859103992708925 \\
7	0.857696149577344 \\
8	0.858712222391874 \\
9	0.85712726324313 \\
10	0.854244961150167 \\
11	0.857508817220528 \\
12	0.857094053154282 \\
13	0.854709907843642 \\
14	0.855874121881354 \\
15	0.85510786797146 \\
16	0.853394178417712 \\
17	0.855131034561963 \\
18	0.856725221169669 \\
19	0.856050802350276 \\
20	0.855161766759616 \\
21	0.857026929938738 \\
22	0.857003395942089 \\
23	0.858666350430549 \\
24	0.855571254263224 \\
};
\addplot [semithick, red, mark=-, mark size=5, mark options={solid}, only marks, forget plot]
table [row sep=\\]{%
0	0.857370988570784 \\
1	0.889989837071926 \\
2	0.897857536515589 \\
3	0.902615146422788 \\
4	0.903599049030277 \\
5	0.901577060806772 \\
6	0.903414352156707 \\
7	0.908273244201654 \\
8	0.907093797343241 \\
9	0.907228283080604 \\
10	0.908851044286626 \\
11	0.90937092351042 \\
12	0.909473433566874 \\
13	0.911041896287881 \\
14	0.911937136210155 \\
15	0.91208772788654 \\
16	0.911409317154316 \\
17	0.910581078374351 \\
18	0.910290850323322 \\
19	0.909101889906635 \\
20	0.909430289783604 \\
21	0.910113561736539 \\
22	0.909703632498501 \\
23	0.911146831777076 \\
24	0.910342849895613 \\
};
\addplot [semithick, color0, solid, mark=*, mark size=2, mark options={solid}, forget plot]
table [row sep=\\]{%
0	0.836044432286252 \\
1	0.869585768981293 \\
2	0.876945760646619 \\
3	0.878348845628502 \\
4	0.878068976748492 \\
5	0.878690300407714 \\
6	0.881259172432816 \\
7	0.882984696889499 \\
8	0.882903009867557 \\
9	0.882177773161867 \\
10	0.881548002718396 \\
11	0.883439870365474 \\
12	0.883283743360578 \\
13	0.882875902065761 \\
14	0.883905629045755 \\
15	0.883597797929 \\
16	0.882401747786014 \\
17	0.882856056468157 \\
18	0.883508035746496 \\
19	0.882576346128455 \\
20	0.88229602827161 \\
21	0.883570245837638 \\
22	0.883353514220295 \\
23	0.884906591103812 \\
24	0.882957052079419 \\
};
\end{axis}

\begin{axis}[
axis x line=top,
height=\figureheight,
tick align=outside,
width=\figurewidth,
x grid style={white!69.01960784313725!black},
xlabel={Average tweets/cascade},
xmin=-1.2, xmax=25.2,
xtick pos=right,
xtick={0,1,2,3,4,5,6,7,8,9,10,11,12,13,14,15,16,17,18,19,20,21,22,23,24},
xticklabels={1,26,31,34,37,38,39,41,42,43,44,44,45,46,47,47,48,48,49,49,50,50,51,51,52},
y grid style={white!69.01960784313725!black},
ymin=0.809849383407478, ymax=0.916956220480781,
ytick pos=left,
ticklabel style = {font=\tiny},
yticklabel=\empty,
x axis line style=-,
]
\end{axis}

\end{tikzpicture}
    \end{minipage}
    \caption{Performance of URL-wise (top) and cascade-wise (bottom) fake news detection (mean ROC AUC, averaged on five folds) as function of cascade diffusion time.}
    \label{fig:diffusion_roc_auc_24h_5_folds}
\end{figure}

Figure~\ref{fig:diffusion_roc_auc_24h_5_folds} depicts the performance of the model (mean ROC AUC) as function of the cascade duration, for the URL- (top) and cascade-wise (bottom)  settings. 
As expected, performance increases with the cascade duration, saturating roughly after 15 hours in the URL-wise setting and after 7 hours in the cascade-wise one, respectively.  
This different behavior is mainly due to the simpler topological patterns and shorter life of individual cascades. 
Seven hours of spreading encompass on average around $91\%$ of the cascade size; for the URL-wise setting, the corresponding value is $68\%$. A similar level of coverage, $86\%$, is achieved after 15 hours of spreading in the URL-wise setting. 

We also note that remarkably just a few ($\sim 2$) hours of news spread are sufficient to achieve above $90\%$ mean ROC AUC in URL-wise fake news classification. 
Furthermore, we observe a significant jump in performance from the 0 hr setting (effectively using only user profile, user activity, and content features) to $\geq1$ hr settings (considering additionally the news propagation), which we interpret as another indication of the importance of propagation-related features.


\subsection{Model aging}

We live in a dynamic world with constantly evolving political context. Since the social network, user preferences and activity, news topics and potentially also spreading patterns evolve in time, it is important to understand to what extent a model trained in the past can generalize to such new circumstances. 
In the final set of experiments, we study how the model performance ages with time in the URL- and cascade-wise settings. 
These experiments aim to emulate a real-world scenario in which a model trained on historical data is applied to new tweets in real time.

For the URL-wise setting, we split our dataset into  training/validation ($80\%$ of URLs) and test  ($20\%$ of URLs) sets; the training/validation and test sets were disjoint and  subsequent in time. 
We assessed the results of our model on subsets of the test set, designed as partially overlapping (mean intersection over union equal to  $0.56\pm0.15$) time windows.  
Partial overlap allowed us to work on larger subsets while preserving the ratio of positives vs negatives, providing at the same time smoother results as with moving average.
This way, each window contained at least $24\%$ of the test set (average number of URLs in a window was $73\pm33.34$) and the average dates of two consecutive windows were at least 14 days apart, progressively increasing.

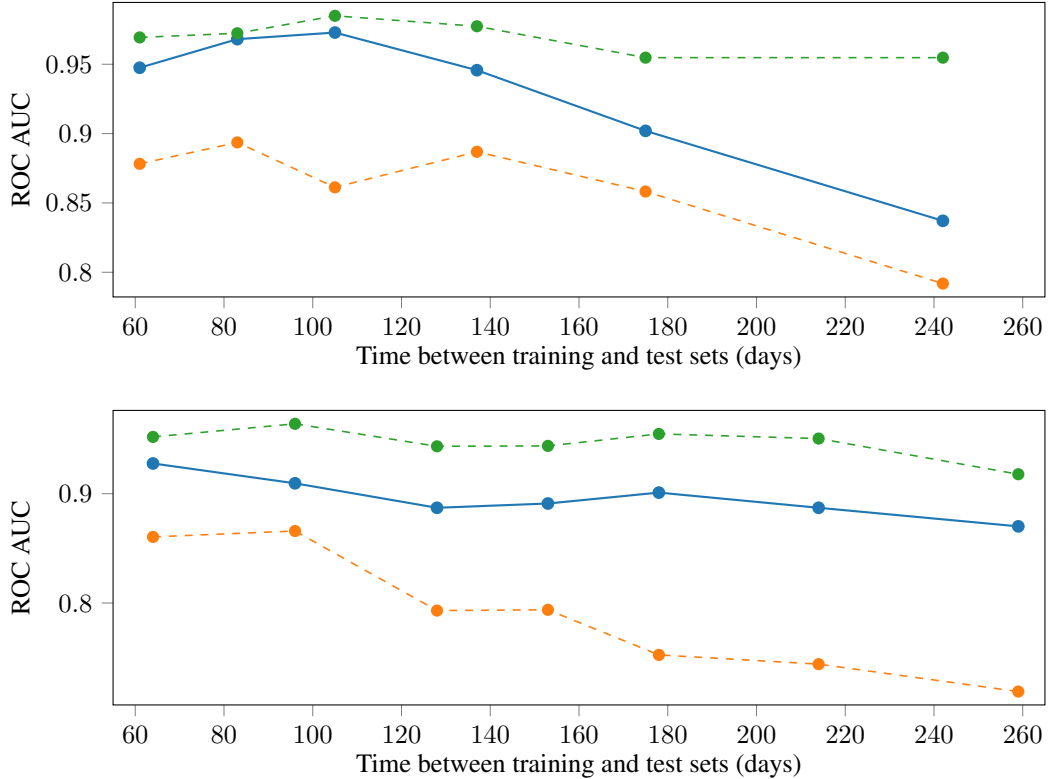
\begin{figure}[t]
    \setlength\figureheight{5.5cm} 
    \setlength\figurewidth{1.0\linewidth}  
    \begin{minipage}{1.0\linewidth}
    \centering
\begin{tikzpicture}

\definecolor{color2}{rgb}{0.172549019607843,0.627450980392157,0.172549019607843}
\definecolor{color0}{rgb}{0.12156862745098,0.466666666666667,0.705882352941177}
\definecolor{color1}{rgb}{1,0.498039215686275,0.0549019607843137}

\begin{axis}[
height=\figureheight,
tick align=outside,
tick pos=left,
width=\figurewidth,
x grid style={white!69.01960784313725!black},
xlabel={Time between training and test sets (days)},
xmin=55, xmax=265,
y grid style={white!69.01960784313725!black},
ylabel={ROC AUC},
ymin=0.782202111613876, ymax=0.994570135746606
]
\addplot [thick, color0, solid, mark=*, mark size=2, mark options={solid}, forget plot]
table [row sep=\\]{%
61	0.947484554280671 \\
83	0.968085106382979 \\
105	0.972850678733032 \\
137	0.945701357466063 \\
175	0.901960784313725 \\
242	0.83710407239819 \\
};
\addplot [semithick, color1, dashed, mark=*, mark size=2, mark options={solid}, forget plot]
table [row sep=\\]{%
61	0.87819947043248 \\
83	0.893617021276596 \\
105	0.861236802413273 \\
137	0.886877828054299 \\
175	0.858220211161388 \\
242	0.791855203619909 \\
};
\addplot [semithick, color2, dashed, mark=*, mark size=2, mark options={solid}, forget plot]
table [row sep=\\]{%
61	0.969329214474846 \\
83	0.972340425531915 \\
105	0.984917043740573 \\
137	0.97737556561086 \\
175	0.95475113122172 \\
242	0.954751131221719 \\
};
\end{axis}

\end{tikzpicture}
    \end{minipage}
    \begin{minipage}{1.0\linewidth}
    \centering
\begin{tikzpicture}

\definecolor{color2}{rgb}{0.172549019607843,0.627450980392157,0.172549019607843}
\definecolor{color0}{rgb}{0.12156862745098,0.466666666666667,0.705882352941177}
\definecolor{color1}{rgb}{1,0.498039215686275,0.0549019607843137}

\begin{axis}[
height=\figureheight,
tick align=outside,
tick pos=left,
width=\figurewidth,
x grid style={lightgray!92.02614379084967!black},
xlabel={Time between training and test sets (days)},
xmin=55, xmax=265,
y grid style={lightgray!92.02614379084967!black},
ylabel={ROC AUC},
ymin=0.706718030247442, ymax=0.976213770331417
]
\addplot [thick, color0, solid, mark=*, mark size=2, mark options={solid}, forget plot]
table [row sep=\\]{%
64	0.927675988428158 \\
96	0.909530583214794 \\
128	0.887122416534181 \\
153	0.891039688336986 \\
178	0.901022644265888 \\
214	0.887122416534181 \\
259	0.870123517182341 \\
};
\addplot [semithick, color1, dashed, mark=*, mark size=2, mark options={solid}, forget plot]
table [row sep=\\]{%
64	0.860495017679203 \\
96	0.865880918512497 \\
128	0.793078146019322 \\
153	0.793766739712686 \\
178	0.752495738982225 \\
214	0.744038155802862 \\
259	0.718967836614895 \\
};
\addplot [semithick, color2, dashed, mark=*, mark size=2, mark options={solid}, forget plot]
table [row sep=\\]{%
64	0.951976856316297 \\
96	0.963963963963964 \\
128	0.943377766907179 \\
153	0.943754565376187 \\
178	0.954711468224982 \\
214	0.950470832823774 \\
259	0.917818270759447 \\
};
\end{axis}

\end{tikzpicture}
    \end{minipage}
    \caption{Effects of training set aging on the performance of 
    URL- (top) and cascade-wise (bottom) fake news detection.  
    Horizontal axis shows difference in days between average date of the training and test sets. Shown is the test performance obtained by our model 
    with 24hrs diffusion (solid blue), test performance obtained with same model just using the first tweet of each piece of news (0hrs diffusion, dashed orange), and test performance obtained training on our original uniformly sampled five folds (veracity predictions are computed for each URL/cascade when this appears as a test sample in our 24hrs five fold cross-validation, green).}
    \label{fig:aging_url}
\end{figure}

Figure \ref{fig:aging_url} (top) captures the variation in performance due to aging of the training set in the URL-wise setting. 
%
Our model exhibits a slight deterioration in performance only after 180 days. We attribute this deterioration to the change in the spreading patterns and the user activity profiles. 

We repeated the same experiment in the cascade-wise setting. 
The split into training/validation and test sets and the generation of the time windows was done  similarly to the URL-wise experiment. 
%
Each time window in the test set has an average size of $314\pm148$ cascades, and two consecutive windows had a mean overlap with intersection over union equal to $0.68\pm0.21$. 
Figure \ref{fig:aging_url} (bottom) summarizes the performance of our model in the cascade-wise setting. 
In this case, it shows a more robust behavior compared to the URL-wise setting, losing only 4\% after 260 days. 


This different behavior is likely due to the higher variability that characterizes cascades as opposed to URLs. As individual cascades are represented by smaller and simpler graphs, the likelihood of identifying recurrent rich structures between different training samples is lower compared to the URL-wise setting and, also, cascades may more easily involve users coming from different parts of the Twitter social network. In the cascade-wise setting, our propagation-based model is thus forced to learn simpler features that on the one hand are less discriminative (hence the lower overall performance), and on the other hand appear to be more robust to aging. %
We leave a deeper analysis of this behavior to future research, which might provide additional ways improving the fake news classification performance. 

 \subsection{Conclusions}
 
 In this paper, we presented a geometric deep learning approach for fake news detection on Twitter social network. The proposed method naturally allows  integrating heterogeneous data pertaining to the user profile and activity, social network structure, news spreading patterns and content. The key advantage of using a deep learning approach as opposed to `handcrafted' features is its ability to automatically learn task-specific features from the data; the choice of geometric deep learning in this case is motivated by the graph-structured nature of the data. 
 Our model achieves very high accuracy and robust behavior in several challenging settings involving large-scale real data, pointing to the great potential of geometric deep learning methods for fake news detection.

 There are multiple intriguing phenomena and hypotheses left for future research. Of particular interest is the experimental validation of the conjecture that our model is potentially language and geography-independent, being mainly based on connectivity and spreading features. The study of adversarial attacks is also of great interest, both from theoretical and practical viewpoints: on the one hand, they adversarial attacks would allow exploration of the limitations of the model and its resilience to attacks. We conjecture that attacks on graph-based approaches require social network manipulations that are difficult to implement in practice, making our method particularly appealing. On the other hand, adversarial techniques could shed light on the way the graph neural network makes decisions, contributing to better interpretability of the model. 
 Finally, we intend to explore additional applications of our model in social network data analysis going beyond fake news detection, such as news topic classification and virality prediction.

\subsubsection*{Acknowledgments}

We gratefully acknowledge the generous support by ERC Consolidator Grant No. 724228 (LEMAN), ERC Proof of Concept grant No. 812672 (GoodNews), multiple Google Research Faculty awards and Nvidia equipment grants, Amazon AWS Machine Learning Research grant, Dalle Molle Foundation prize, and Facebook Computational Social Science Methodology award. 
MB is also partially supported by the Royal Society Wolfson Research Merit award and Rudolf Diesel industrial fellowship at TU Munich. 
FF's PhD scholarship is supported by the SNF Grant No. 200021E/176315. 
%


{\small
\bibliographystyle{plain}
\bibliography{sections/biblio.bib}

\begin{thebibliography}{10}

\bibitem{afroz2012detecting}
Sadia Afroz, Michael Brennan, and Rachel Greenstadt.
\newblock Detecting hoaxes, frauds, and deception in writing style online.
\newblock In {\em Proc. IEEE Symp. Security and Privacy (SP)}, pages 461--475,
  2012.

\bibitem{boscaini2016learning}
Davide Boscaini, Jonathan Masci, Emanuele Rodol{\`a}, and Michael Bronstein.
\newblock Learning shape correspondence with anisotropic convolutional neural
  networks.
\newblock In {\em Proc. NIPS}, 2016.

\bibitem{bovet2019influence}
Alexandre Bovet and Hern{\'a}n~A Makse.
\newblock Influence of fake news in {T}witter during the 2016 {US} presidential
  election.
\newblock {\em Nature Communications}, 10(1):7, 2019.

\bibitem{bronstein2017geometric}
Michael~M Bronstein, Joan Bruna, Yann LeCun, Arthur Szlam, and Pierre
  Vandergheynst.
\newblock Geometric deep learning: going beyond euclidean data.
\newblock {\em IEEE Signal Processing Magazine}, 34(4):18--42, 2017.

\bibitem{bruna2014spectral}
Joan Bruna, Wojciech Zaremba, Arthur Szlam, and Yann Lecun.
\newblock Spectral networks and locally connected networks on graphs.
\newblock In {\em Proc. ICLR}, 2014.

\bibitem{choma2018graph}
Nicholas Choma, Federico Monti, Lisa Gerhardt, Tomasz Palczewski, Zahra
  Ronaghi, Prabhat Prabhat, Wahid Bhimji, Michael Bronstein, Spencer Klein, and
  Joan Bruna.
\newblock Graph neural networks for icecube signal classification.
\newblock In {\em Proc. ICMLA}, 2018.

\bibitem{conover2011political}
Michael Conover, Jacob Ratkiewicz, Matthew~R Francisco, Bruno Gon{\c{c}}alves,
  Filippo Menczer, and Alessandro Flammini.
\newblock Political polarization on twitter.
\newblock In {\em Proc. ICWSM}, 2011.

\bibitem{defferrard2016convolutional}
Micha{\"e}l Defferrard, Xavier Bresson, and Pierre Vandergheynst.
\newblock Convolutional neural networks on graphs with fast localized spectral
  filtering.
\newblock In {\em Proc. NIPS}, 2016.

\bibitem{fruchterman1991graph}
Thomas~MJ Fruchterman and Edward~M Reingold.
\newblock Graph drawing by force-directed placement.
\newblock {\em Software: Practice and experience}, 21(11):1129--1164, 1991.

\bibitem{gilmer2017neural}
Justin Gilmer, Samuel~S Schoenholz, Patrick~F Riley, Oriol Vinyals, and
  George~E Dahl.
\newblock Neural message passing for quantum chemistry.
\newblock In {\em Proc. ICML}, 2017.

\bibitem{howell2013digital}
Lee Howell et~al.
\newblock Digital wildfires in a hyperconnected world.
\newblock {\em WEF Report}, 3:15--94, 2013.

\bibitem{kipf2016semi}
Thomas~N Kipf and Max Welling.
\newblock Semi-supervised classification with graph convolutional networks.
\newblock 2017.

\bibitem{klambauer2017self}
G{\"u}nter Klambauer, Thomas Unterthiner, Andreas Mayr, and Sepp Hochreiter.
\newblock Self-normalizing neural networks.
\newblock In {\em Proc. NIPS}, 2017.

\bibitem{kucharski2016post}
Adam Kucharski.
\newblock Post-truth: Study epidemiology of fake news.
\newblock {\em Nature}, 540(7634):525, 2016.

\bibitem{kwon2013prominent}
Sejeong Kwon, Meeyoung Cha, Kyomin Jung, Wei Chen, and Yajun Wang.
\newblock Prominent features of rumor propagation in online social media.
\newblock In {\em Proc. Conf. Data Mining}, pages 1103--1108, 2013.

\bibitem{lazer2018science}
David~MJ Lazer, Matthew~A Baum, Yochai Benkler, Adam~J Berinsky, Kelly~M
  Greenhill, Filippo Menczer, Miriam~J Metzger, Brendan Nyhan, Gordon
  Pennycook, David Rothschild, et~al.
\newblock The science of fake news.
\newblock {\em Science}, 359(6380):1094--1096, 2018.

\bibitem{lecun2015deep}
Yann LeCun, Yoshua Bengio, and Geoffrey Hinton.
\newblock Deep learning.
\newblock {\em nature}, 521(7553):436, 2015.

\bibitem{lecun1998gradient}
Yann LeCun, L{\'e}on Bottou, Yoshua Bengio, and Patrick Haffner.
\newblock Gradient-based learning applied to document recognition.
\newblock {\em Proc. IEEE}, 86(11):2278--2324, 1998.

\bibitem{levie2017cayleynets}
Ron Levie, Federico Monti, Xavier Bresson, and Michael~M Bronstein.
\newblock Cayleynets: Graph convolutional neural networks with complex rational
  spectral filters.
\newblock {\em arXiv:1705.07664}, 2017.

\bibitem{litany2018deformable}
Or~Litany, Alex Bronstein, Michael Bronstein, and Ameesh Makadia.
\newblock Deformable shape completion with graph convolutional autoencoders.
\newblock In {\em Proc. CVPR}, 2018.

\bibitem{long2017fake}
Yunfei Long, Qin Lu, Rong Xiang, Minglei Li, and Chu-Ren Huang.
\newblock Fake news detection through multi-perspective speaker profiles.
\newblock In {\em Proc. Natural Language Processing}, volume~2, pages 252--256,
  2017.

\bibitem{masci2015geodesic}
Jonathan Masci, Davide Boscaini, Michael Bronstein, and Pierre Vandergheynst.
\newblock Geodesic convolutional neural networks on riemannian manifolds.
\newblock In {\em Proc. ICCV Workshops}, 2015.

\bibitem{monti2016geometric}
Federico Monti, Davide Boscaini, Jonathan Masci, Emanuele Rodol{\`a}, Jan
  Svoboda, and Michael~M Bronstein.
\newblock Geometric deep learning on graphs and manifolds using mixture model
  {CNN}s.
\newblock In {\em Proc. CVPR}, 2017.

\bibitem{monti2017geometric}
Federico Monti, Michael Bronstein, and Xavier Bresson.
\newblock Geometric matrix completion with recurrent multi-graph neural
  networks.
\newblock In {\em Proc. NIPS}, 2017.

\bibitem{monti2018motifnet}
Federico Monti, Karl Otness, and Michael~M Bronstein.
\newblock Motifnet: a motif-based graph convolutional network for directed
  graphs.
\newblock In {\em Proc. Data Science Workshop}, 2018.

\bibitem{monti2018dual}
Federico Monti, Oleksandr Shchur, Aleksandar Bojchevski, Or~Litany, Stephan
  G{\"u}nnemann, and Michael~M Bronstein.
\newblock Dual-primal graph convolutional networks.
\newblock {\em arXiv:1806.00770}, 2018.

\bibitem{pennington2014glove}
Jeffrey Pennington, Richard Socher, and Christopher Manning.
\newblock Glove: Global vectors for word representation.
\newblock In {\em Proc. {EMNLP}}, 2014.

\bibitem{perez2017automatic}
Ver{\'o}nica P{\'e}rez-Rosas, Bennett Kleinberg, Alexandra Lefevre, and Rada
  Mihalcea.
\newblock Automatic detection of fake news.
\newblock {\em arXivarXiv:1708.07104}, 2017.

\bibitem{potthast2017stylometric}
Martin Potthast, Johannes Kiesel, Kevin Reinartz, Janek Bevendorff, and Benno
  Stein.
\newblock A stylometric inquiry into hyperpartisan and fake news.
\newblock {\em arXiv:1702.05638}, 2017.

\bibitem{rashkin2017truth}
Hannah Rashkin, Eunsol Choi, Jin~Yea Jang, Svitlana Volkova, and Yejin Choi.
\newblock Truth of varying shades: Analyzing language in fake news and
  political fact-checking.
\newblock In {\em Proc. Empirical Methods in Natural Language Processing},
  pages 2931--2937, 2017.

\bibitem{reddi2018convergence}
Sashank~J Reddi, Satyen Kale, and Sanjiv Kumar.
\newblock On the convergence of {ADAM} and beyond.
\newblock 2018.

\bibitem{rubin2016fake}
Victoria Rubin, Niall Conroy, Yimin Chen, and Sarah Cornwell.
\newblock Fake news or truth? using satirical cues to detect potentially
  misleading news.
\newblock In {\em Proc. Computational Approaches to Deception Detection}, pages
  7--17, 2016.

\bibitem{ruchansky2017csi}
Natali Ruchansky, Sungyong Seo, and Yan Liu.
\newblock Csi: A hybrid deep model for fake news.
\newblock {\em arXiv:1703.06959}, 2017.

\bibitem{scarselli2009graph}
Franco Scarselli, Marco Gori, Ah~Chung Tsoi, Markus Hagenbuchner, and Gabriele
  Monfardini.
\newblock The graph neural network model.
\newblock {\em IEEE Trans. Neural Networks}, 20(1):61--80, 2009.

\bibitem{shu2019studying}
Kai Shu, H~Russell Bernard, and Huan Liu.
\newblock Studying fake news via network analysis: detection and mitigation.
\newblock In {\em Emerging Research Challenges and Opportunities in
  Computational Social Network Analysis and Mining}, pages 43--65. Springer,
  2019.

\bibitem{shu2017fake}
Kai Shu, Amy Sliva, Suhang Wang, Jiliang Tang, and Huan Liu.
\newblock Fake news detection on social media: A data mining perspective.
\newblock {\em ACM SIGKDD Explorations Newsletter}, 19(1):22--36, 2017.

\bibitem{shu2018understanding}
Kai Shu, Suhang Wang, and Huan Liu.
\newblock Understanding user profiles on social media for fake news detection.
\newblock In {\em Proc. Multimedia Information Processing and Retrieval}, pages
  430--435, 2018.

\bibitem{shu2019beyond}
Kai Shu, Suhang Wang, and Huan Liu.
\newblock Beyond news contents: The role of social context for fake news
  detection.
\newblock In {\em Proc. Web Search and Data Mining}, 2019.

\bibitem{svoboda2018peernets}
Jan Svoboda, Jonathan Masci, Federico Monti, Michael~M Bronstein, and Leonidas
  Guibas.
\newblock Peernets: Exploiting peer wisdom against adversarial attacks.
\newblock In {\em Proc. ICLR}, 2019.

\bibitem{tacchini2017some}
Eugenio Tacchini, Gabriele Ballarin, Marco~L Della~Vedova, Stefano Moret, and
  Luca de~Alfaro.
\newblock Some like it hoax: Automated fake news detection in social networks.
\newblock {\em arXiv:1704.07506}, 2017.

\bibitem{velickovic2017graph}
Petar Velickovic, Guillem Cucurull, Arantxa Casanova, Adriana Romero, Pietro
  Lio, and Yoshua Bengio.
\newblock Graph attention networks.
\newblock In {\em Proc. ICLR}, 2018.

\bibitem{vosoughi2018spread}
Soroush Vosoughi, Deb Roy, and Sinan Aral.
\newblock The spread of true and false news online.
\newblock {\em Science}, 359(6380):1146--1151, 2018.

\bibitem{wang2018dynamic}
Yue Wang, Yongbin Sun, Ziwei Liu, Sanjay~E Sarma, Michael~M Bronstein, and
  Justin~M Solomon.
\newblock Dynamic graph cnn for learning on point clouds.
\newblock {\em arXiv:1801.07829}, 2018.

\bibitem{zhou2018fake}
Xinyi Zhou and Reza Zafarani.
\newblock Fake news: A survey of research, detection methods, and
  opportunities.
\newblock {\em arXiv:1812.00315}, 2018.

\end{thebibliography}
}

\end{document}